\documentclass{article}

\usepackage[version=4]{mhchem}
\usepackage{tipx}
\usepackage{threeparttable}
\usepackage[margin=1in]{geometry} 
\usepackage{graphicx}
\usepackage{color}
\usepackage[sort&compress,numbers,square]{natbib}

\title{Mechanistic insights into water autoionization}
\author{Ling Liu$^{1,2}$ \and Yingqi Tian$^{1,2}$ \and Chungen Liu$^{1,2*}$}

\date{$^1$Institute of Theoretical and Computational Chemistry, School of Chemistry and Chemical Engineering, Nanjing University, Nanjing, 210023, China \\
     $^2$Jiangsu Key Laboratory of Vehicle Emissions Control, Nanjing University, Nanjing 210023, China \\
	$^*$cgliu@nju.edu.cn
}

\begin{document}
	
\maketitle

\begin{abstract}
 
Water autoionization plays a critical role in determining pH and properties of various chemical and biological processes occurring in the water mediated environment. The strikingly unsymmetrical potential energy surface of the dissociation process poses a great challenge to the mechanistic study. Here, we demonstrate that reliable sampling of the ionization path is accessible through nanosecond timescale metadynamics simulation enhanced by machine learning of the neural network potentials with ab initio precision, which is proved by quantitatively reproduced water equilibrium constant (p$K_\mathrm{w}$=14.14) and ionization rate constant (1.566$\times10^{-3}$ s$^{-1}$). Statistical analysis unveils the asynchronous character of the concerted triple proton transfer process. Based on conditional ensemble average calculations, we propose a dual-presolvation mechanism, which suggests that a pair of hypercoordinated and undercoordinated waters bridged by one \ce{H2O} cooperatively constitutes the initiation environment for autoionization, and contributes majorly to the local electric field fluctuation to promote water dissociation.
 
 
\noindent \textbf{keywords:} water autoionization, atomic neural network potentials, metadynamics, multiple proton transfer, presolvation
\end{abstract}

\maketitle

\section{Introduction}\label{sec1}

Water serves as the most basic and common substance involved in a wide variety of chemical processes and living organisms \cite{Marx2010,Agmon2016,Ceriotti2016}. However, the hydrogen bonds (HBs) formed in liquid water make it access to many anomalous properties \cite{Ceriotti2013,Morawietz2016,Chen2018}. Among them, water autoionization plays a critical role in determining pH and conductivity, where the active water wire along with its surrounding solvents undergoes several breakage and reconstitution of hydrogen bonds. Besides the difficulty in monitoring the dynamically structured hydrogen bond network, the multiscale kinetics feature of water ionization arising from the striking stability difference between water and dissociated ions \cite{Eigen1958,Natzle1985} also challenges the interpretation of autoionization mechanism. 

Decades ago, Eigen and De Mayer proposed that the initial dissociation of water system is the dynamical bottleneck for autoionization process, where the formed \ce{H3O+} and \ce{OH-} ions reach the so-called contact distance, followed by a relatively facile diffusion away from each other~\cite{Eigen1958}. The entire process can be expressed as follows, 
\begin{equation}
	\ce{H2O} \ce{<=>[$k_\mathrm{d1}$][$k_\mathrm{r1}$]} \ce{H3O+}\ce{\bond{...}}\ce{OH-} \ce{<=>[$k_\mathrm{d2}$][$k_\mathrm{r2}$]} \ce{H3O+}+\ce{OH-}	
	\label{eq:diss}
\end{equation}
The above kinetic model was later employed by Natzle et al. in an experimental study of the transient conductivity in photoionized water system \cite{Natzle1985}. Based on Eigen's assumption that the reverse recombination rate is controlled by diffusion of free ions ($k_\mathrm{r1}\gg k_\mathrm{r2}$), the contact distance was characterized as 5.8 $\pm$ 0.5 \AA, corresponding to an ion-pair separation by two waters. Meanwhile, the overall recombination rate constant, approximately equal to $k_\mathrm{r2}$, was determined as 0.112 M$^{-1}$$\cdot$ps$^{-1}$, and the dissociation rate constant was estimated as $2.04\times10^{-5}$ s$^{-1}$ by taking the well-characterized ionization equilibrium constant of water. Due to the extremely scarce probability of water dissociation in natural conditions, directly tracking the autoionization process of a specific water molecule still remains a tough task today. 

Given the rare-event character of water ionization, theoretical study has demonstrated its irreplaceable role to complement experimental exploration of the mechanism \cite{Trout1998,Sprik2000,Geissler2001,Hassanali2011,Moqadam2018}. Earlier studies employed constrained ab initio molecular dynamics (AIMD) to compute the free energy profile along the chosen reaction coordinate \cite{Trout1998,Sprik2000}, which was predefined as the stretching of a specific O-H bond, or evolution of the number of covalent bonds on a selected oxygen. These structural constraints are lack of adjustability to fully describe the whole dissociation progress, which is proved to experience a series of chemical bond reconstitution. Introducing the path-sampling methods into AIMD provides another solution to simulate rare-event processes that is free of the inconvenience of defining geometrical constraints \cite{Dellago1998}. After analyzing very limited number of transition paths, Geissler et al. revealed that the water dissociation can be substantially accelerated by the solvent electric field \cite{Geissler2001,Geissler2005}. The reverse neutralization process is kinetically fast and can be theoretically explored with standard AIMD \cite{Hassanali2011}. By inspecting the neutralization features from hundreds of AIMD trajectories, Hassanali et al. deduced that the water ionization process goes through a concerted triple proton jump, which is triggered by a collective compression of the reactive four-water wire and the presolvation phenomenon characterized as the hypercoordination of the dissociating water. These initiation conditions for water ionization were essentially confirmed by Moqadam et al. in a more recent study, by performing AIMD with replica exchange transition interface sampling (RETIS) method \cite{Moqadam2018}. However, they proposed that the dissociation event is likely to occur through a double proton jump, leaving the third proton transfer in a stepwise way. This view seems to contradict the triple proton jump mechanism from the neutralization simulation \cite{Moqadam2018}.  


The pioneering theoretical explorations contributed significantly in establishing the fundamental atomic-level mechanism for water autoionization, although there remains some essential problems. Firstly, it is known that the construction of a set of reaction coordinates is an essential prerequisite to enable the rational description of ionization progress in the free energy landscapes. However, the artificially chosen coordinates in the constrained AIMD \cite{Trout1998,Sprik2000} cannot match up with the latest reaction mechanism, and thus fail to characterize the critical intermediate of ion pair as well as the transition state. Secondly, the dispute on mechanism of proton transfer in the water ionization \cite{Hassanali2011,Moqadam2018} may attribute to the lack of guarantee for equilibrium sampling of structural distributions along the whole reaction pathway due to the limited number of AIMD trajectories, which is expected to be clarified by characterizing a converged free energy path for the formation of ion pair that separated by two waters. As a consequence, the presolvation phenomenon and the electric field fluctuation should also be revisited based on the construction of a reliable statistical ensemble which can properly reflect the equilibrium probabilities of both normal and rare states.

In this work, we employ the well-tempered metadynamics method to characterize the free energy profiles of water autoionization by iteratively ``filling'' the potential energy of the system \cite{Laio2002,Parrinello2008}, which was previously proved to well describe the typical acid-base reactions based on introducing good collective variables (CVs) \cite{Grifoni2019}. Here a series of CVs is specifically designed to monitor the proton transfer progress and the evolution of concerned properties from the metadynamics trajectory. Considering the deep potential well in the neutral water state, a nanosecond (ns) timescale is required to ensure the convergence of free energy profiles and hence the sufficient statistical sampling of ionization path. To overcome the computational bottleneck of ab initio metadynamics simulation, atomic neural network potentials (NNPs) are employed to replace the time-consuming density functional theory (DFT) calculation, which was reported to accurately represent the ab initio potential energy and atomic forces with a low computational demand comparable to the classical force field methods \cite{Behler2007,Behler2011}. We will show that the equilibrium constant and rate constant of water autoionization deduced from the simulated free energy profile are highly consistent with the documented experimental value. By looking into the structural evolution along the CVs, we will reveal the asynchronous nature of concerted triple proton transfer along the hydrogen-bonded water wire, and clarify the push-pull effect arising from the dual presolvation which contributes to the electric field fluctuations on triggering the water ionization.

\section{Results and Discussion}\label{sec2}

\begin{figure}
	\centering
	\includegraphics[width=0.7\linewidth] {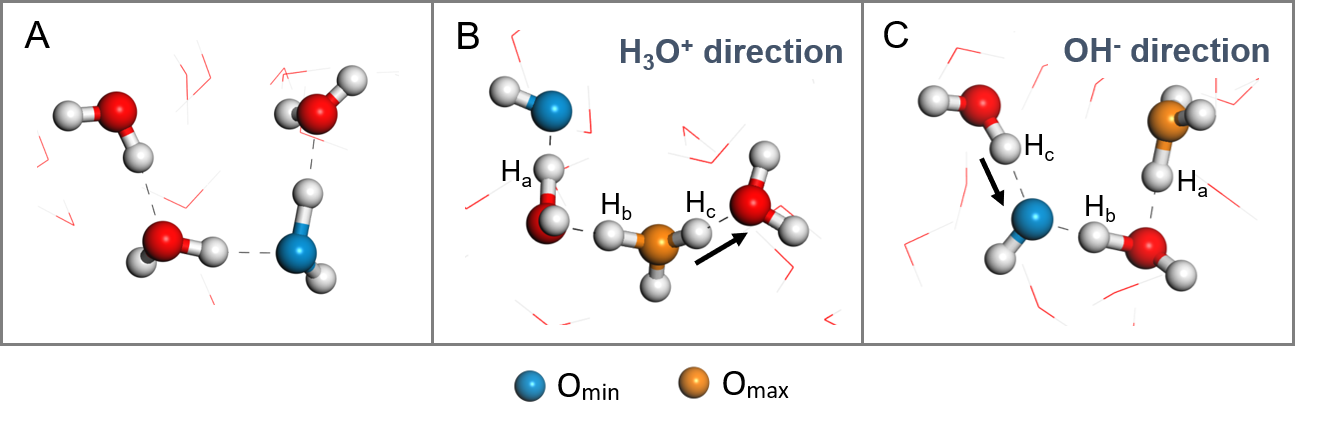}
	\caption{Schematic diagram of four-water wires selected for exploring the water ionization process. The O centers of water, hydronium or hydroxide ions that possess the utmost negative and positive charges are marked with O$_\text{min}$ and O$_\text{max}$, respectively. (A) A selected water wire containing only O$_\text{min}$ in a snapshot of neutral water state. (B) and (C) are the selected water wires containing O$_\text{min}$ and O$_\text{max}$ in the snapshots of ion-pair state, in which the proton transfer proceeds along \ce{H3O+} and \ce{OH-} directions, respectively. The three bridging hydrogen atoms along the hydrogen-bonded wire are labeled H$_\mathrm{a}$, H$_\mathrm{b}$ and H$_\mathrm{c}$, according to the sequence of proton transfer progress (by comparing the elongated O-H bond lengths), and H$_\mathrm{c}$ is ensured at the side. 	}
	\label{fig:hbwire}
\end{figure}


\subsection{Ionization free energy profiles}

In performing metadynamics simulations, we introduce two CVs for controlling the progress of water dissociation, which relate to the number of ions in the system and the distance between ionic species (see Equation \ref{eq:cv1} and \ref{eq:cv2} below). Three independent NNPs-based metadynamics with varying Gaussian deposition parameters are performed to examine the consistence of free energy profiles. Up to 10 ns of simulation time is proved to guarantee the convergence of free energy differences and thus the proper statistical sampling along the concerned reaction pathway (see Supplementary Fig. S2). Computational details can be found in Methods Section. We focus on analyzing the early stages of water ionization events, during which one covalent \ce{O-H} is broken, and concomitant with the intermolecular proton transfers (PTs), to create a \ce{H3O+}\ce{\bond{...}}\ce{OH-} pair at the contact distance. Thus, modeling of the ionic pair generation progress should be implemented on the hydrogen-bonded wires comprising at least four waters. Accordingly, we base on the four-water molecule wires to construct a series of CVs for the following discussions. Although the practical scheme of determining the molecule wires for each configuration in the metadynamics trajectory is a bit too complicated as shown in Supplementary Fig. S3, the physical idea of the selection rules can still be expressed concisely. As illustrated in Figure \ref{fig:hbwire}A, the hydrogen-bonded wire in the neutral state snapshot is selected as the shortest one among those which contain the utmost charged O center (O$_\text{min}$ or O$_\text{max}$, determined by $q_i$ in Equation \ref{eq:cv2}). While in the ion-pair state snapshot, the wire is selected as the shortest one that holds the ionic pair. It is worth mentioning that the proton transfers concurrent with the ions formation should be distinguished as along \ce{H3O+} (Figure \ref{fig:hbwire}B) and \ce{OH-} (Figure \ref{fig:hbwire}C) directions based on the subsequent proton transfer mode (black arrows), although they are merged in part of our discussions.

\begin{figure}
	\centering
	\includegraphics[width=0.7\linewidth] {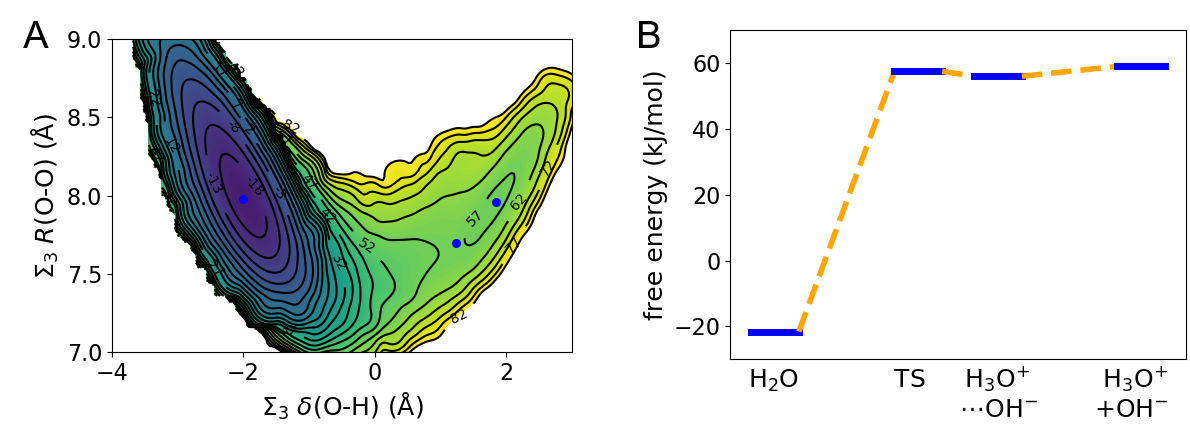}
	\caption{Free energy profiles for water ionization. (A) The 2D free energy surface constructed with two custom CVs. $\sum_3\delta$(O-H) is the sum of all three $\delta$(O-H) for characterizing the overall transfer progress of the three protons. Each $\delta$(O-H) denotes the difference in distances between the moving proton and its neighboring oxygen atoms. If $\delta$(O-H)$>$0, the H is biased towards the target O. $\sum_3 R$(O-O) is the sum of all three $R$(O-O) for measuring the extent of wire contraction, and each $R$(O-O) denotes the pairwise oxygen distance. Free energy values labeled in the contour plot are in units of kJ/mol (the same below). (B) The free energy diagram of critical points, including neutral water state, transition state (TS), ion-pair intermediate (\ce{H3O+}\ce{\bond{...}}\ce{OH-}), and the final well-separated ionic state (\ce{H3O+}+\ce{OH-}). The free energy of final state is evaluated by applying the electrostatic energy correction on the ion-pair state. }
	\label{fig:fes}
\end{figure}

The ionization free energy profiles are computed by reweighting the biased probability density \cite{Tiwary2015} as a function of the chosen CVs, $\sum_3 \delta$(O-H) and $\sum_3 R$(O-O), which quantify the collective behaviors of proton and oxygen atoms within the selected water wires, respectively. It provides a clear demonstration of the reaction path when combined with nudged elastic band (NEB) analysis \cite{Henkelman2008}, as is implemented with fast inertial relaxation engine (FIRE) optimizer \cite{Bitzek2006} to seek the minimum free energy path and transition state on the two-dimensional (2D) CV space. As seen in Figure \ref{fig:fes}A, the overall transfer progress of the involved three protons is strongly coupled with the compression of the whole water wire, showing good consistency with the previous study of \ce{H3O+} and \ce{OH-} recombination process \cite{Hassanali2011}. It can be concluded that the collective compression of water wire plays a determinative role in facilitating the water dissociation event. The unique transition state (TS) located in the 2D free energy surface indicates a concerted triple proton transfer within the same elementary reaction step. The corresponding free energy barrier can be easily determined by extracting the free energies of critical points along the NEB path, as shown in Figure \ref{fig:fes}B. As an average of the three independent metadynamics simulations (see Supplementary Fig. S2 for details), the forward/dissociation and backward/recombination activation free energies, $\Delta G^{\text{\textdoublebarpipevar}}_\mathrm{D}$ and $\Delta G^{\text{\textdoublebarpipevar}}_\mathrm{R}$, are calculated to be 79.363 $\pm$ 0.693 and 1.688 $\pm$ 0.229 kJ/mol, respectively, where a correction of +0.349 kJ/mol has been added to the free energies of both TS and ion-pair state for setting 1 M concentration as the reference.  


We estimate the rate constants by applying variational transition state theory (VTST) \cite{Truhlar1983,Truhlar2017}, expressed as the following formula, 
\begin{equation}
	k=\frac{k_\mathrm{B}T}{h}\exp\left(-\frac{\Delta G^{\text{\textdoublebarpipevar}}}{RT}\right)	
\end{equation}

\noindent  Where $k_\mathrm{B}$, $h$ and $R$ are Boltzmann constant, Planck constant and gas constant, respectively. The calculated elementary rate constants for water dissociation ($k_\mathrm{d1}$) to form the ion-pair intermediate and the backward neutralization ($k_\mathrm{r1}$) are listed in Table \ref{tb:k}. The calculated $k_\mathrm{d1}$ is on the same scale with 0.011 s$^{-1}$ evaluated from the RETIS simulation where a time threshold of 1 ps was set before the ion pair recombines \cite{Moqadam2018}. The ion pair at the contact distance recombines with a frequency of approximately 2.73 ps$^{-1}$ when taking into account the ionic concentration ($\sim$0.869 M) in the simulated box, which shows a satisfactory agreement with the timescale of $\sim$0.5 ps for the collective compression of water wire \cite{Hassanali2011}. The equilibrium constant of water autoionization ($K_\mathrm{w}$) can be feasibly obtained, after introducing the electrostatic energy correction for further ion-pair separation from the contact distance $R_\mathrm{cd}$ to infinity (the isolated state), which can be estimated as the negative of Coulomb interaction energy between two point charges,

\begin{equation}
\Delta E_\mathrm{el}=-\frac{1}{4\pi \varepsilon_0 \varepsilon_r}\frac{q_{\ce{H3O+}}q_{\ce{OH-}}}{R_\mathrm{cd}}=2.993 ~\text{kJ/mol}	
\end{equation}

\noindent Where $\varepsilon_0$ denotes vacuum permittivity, and $\varepsilon_r$ is relative permittivity of water (78.3). The electric charges of \ce{H3O+} and \ce{OH-} are set to +1e and -1e, respectively. The contact distance $R_\mathrm{cd}$ is calculated to be 5.928 \AA~with the ensemble average algorithm described in Methods Section, which agrees well with the experimental data \cite{Natzle1985}. With the electrostatic energy correction $\Delta E_\mathrm{el}$, the value of $K_\mathrm{w}$ is calculated with

\begin{equation}
	K_\mathrm{w}=\frac{k_\mathrm{d1}}{k_\mathrm{r1}}\exp\left(-\frac{\Delta E_\mathrm{el}}{RT}\right)
\end{equation}

\noindent $K_\mathrm{w}$ is dimensionless since the concentration unit in all rate constants is removed by utilizing $c^{\ominus}$, the standard concentration equal to 1 M. The calculated $K_\mathrm{w}$ of $0.724\times10^{-14}$ (p$K_\mathrm{w}$=14.14) is quite close to the well-known experimental data (Table \ref{tb:k}), and shows a higher accuracy than the calculated 13 and 13.7 based on constrained AIMD \cite{Trout1998,Sprik2000} and a recent umbrella sampling simulation \cite{Joutsuka2022}, respectively. Indeed, the identification of local minimum state and transition state is quite ambiguous in these studies which mainly ascribes to the inappropriately chosen reaction coordinates. The quantitative agreement in this work could raise strong confidence in the reliability of the DFT-parameterized neural network potentials and the metadynamics computational schemes, which enables a good description as well as sufficient statistical sampling of the neutral water and ionic states. 

\begin{table}[h]
	\centering
	\caption{Comparison of theoretical and experimental rate constants and equilibrium constant for water ionization at room temperature (298 K). The standard deviations are computed according to the propagation of uncertainty rules.  }
	\begin{tabular}{lcc}
		\hline
		& This work & Experiment \cite{Natzle1985} \\ 
		\hline
		$k_\mathrm{d1}$ (s$^{-1}$) & $0.076\pm0.021$ &  - \\
		$k_\mathrm{d2}$ (ps$^{-1}$) & - & 0.0661 \\
		$k_\mathrm{r1}$ (ps$^{-1}$) & $3.142\pm0.291$ & - \\
		$k_\mathrm{r2}$ (ps$^{-1}$) & - & $0.112$ \\ 
		$k_\mathrm{D}$ (s$^{-1}$)  & $1.566\times10^{-3}$ & $1.132\times10^{-3}$$^*$ \\
		$k_\mathrm{R}$ (ps$^{-1}$)  & $0.216$$^{**}$ & $0.112$ \\ 
		$K_\mathrm{w}$ & $(0.724\pm0.213)\times10^{-14}$ & $ 1\times 10^{-14}$ \\ 
		\hline
	\end{tabular}

	\footnotesize{$^*$The original data $2.04\times10^{-5}$ s$^{-1}$ deduced from $K_\mathrm{w}k_\mathrm{R}/c_{\ce{H2O}}$ has been multiplied with the water concentration of 55.5 M. \\} 
	\footnotesize{$^{**}$$k_\mathrm{R}$ is calculated with $k_\mathrm{D}/K_\mathrm{w}$.} 
	\label{tb:k}
\end{table}


By inspecting Table \ref{tb:k}, it is found that the water dissociation step is rate-determining in the overall autoionization process, since the rate constant $k_\mathrm{d1}$ is twelve orders of magnitude lower than that of the diffusion step of ion pair ($k_\mathrm{d2}$). Accordingly, it is reasonable to assume that the concentration of the intermediate state \ce{H3O+}\ce{\bond{...}}\ce{OH-} is much smaller compared to those of the ionic species and pure water, such that the overall ionization rate constant $k_\mathrm{D}$ can be deduced from the steady-state approximation \cite{Natzle1985}. Then $k_\mathrm{D}$ is calculated with the expression $k_\mathrm{d1}k_\mathrm{d2}/(k_\mathrm{d2}+k_\mathrm{r1})$, with the diffusion rate constants, $k_\mathrm{d2}$ and $k_\mathrm{r2}$, directly from the experimental measurement \cite{Natzle1985}. It should be noted that the calculated $k_\mathrm{D}$ has incorporated the molar concentration of water ($c_{\ce{H2O}}$), thus a correction of experimental $k_\mathrm{D}$ is made for unifying the definition. As listed in Table \ref{tb:k}, the calculated $k_\mathrm{D}$ shows quantitative agreement with the experimental result. This good consistence of rate constants between theoretical simulation and experiment indicates that the assumption made by Eigen where the recombination process is diffusion controlled \cite{Eigen1958} can be reasonably applied in the water ionization system.

\begin{figure}[h]
	\centering
	\includegraphics[width=1.0\linewidth] {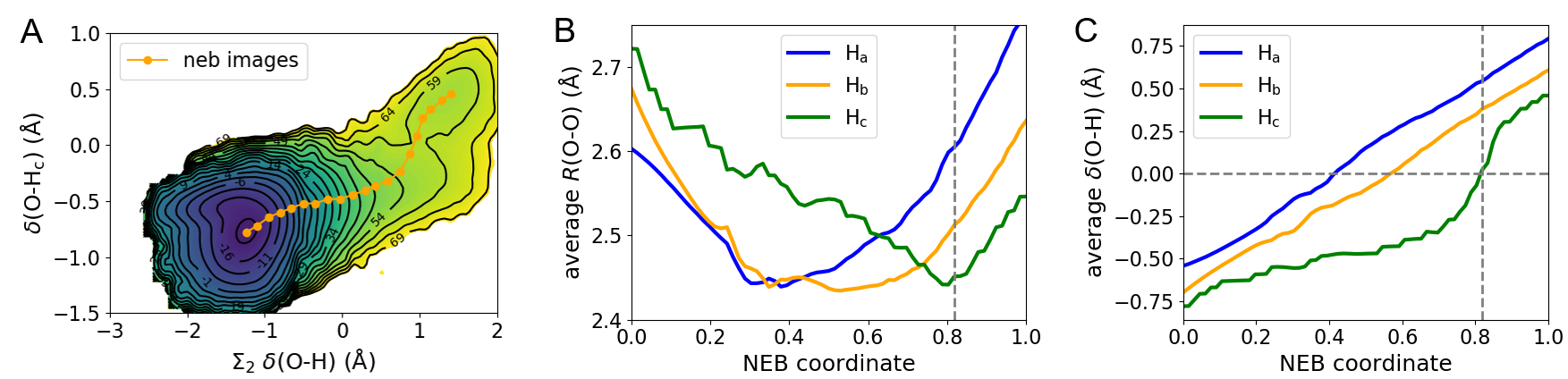}
	\caption{Oxygen and hydrogen motions during water dissociation. (A) The free energy surface constructed with $\delta$(O-H$_\mathrm{c}$) and the sum of $\delta$(O-H$_\mathrm{a}$) and $\delta$(O-H$_\mathrm{b}$). The reaction path shown with connected orange dots is determined by NEB method. Linear interpolation is applied to project the NEB path onto 2D CV grids. The evolution of conditional ensemble average of (B) $R$(O-O) and (C) $\delta$(O-H) along the normalized NEB coordinates. The legend labels the type of H that is in between the two concerned O atoms. The gray vertical line marks the TS location, and the horizontal one highlights the position where $\delta$(O-H) is equal to zero. }
	\label{fig:prop}
\end{figure}

\subsection{Asynchronous proton transfers}

In order to assess the synchronicity of triple proton jump during water dissociation, we decompose one of the CVs in Figure \ref{fig:fes}A, $\sum_3 \delta$(O-H), into $\delta$(O-H$_\mathrm{c}$) and $\delta$(O-H$_\mathrm{a}$)+$\delta$(O-H$_\mathrm{b}$). A new contour map of free energy surface, shown in Figure \ref{fig:prop}A, is then constructed with two redefined CVs, while integrating out $\sum_3 R$(O-O) in recounting the distribution of the biased probability density. We note that the activation free energies obtained here present small differences (within 2 kJ/mol) from that in Figure \ref{fig:fes}A, indicating the redefined CVs can still give a clear distinction among the critical states \cite{Bussi2020}.  
Following the definition by Dewar, a concerted reaction means a reaction takes place in a single kinetic step, and the primitive changes concerned (generally bond rupture and bond formation) are not necessarily fulfilled synchronously \cite{Dewar1984}. However, a synchronous reaction process has more strict requirements, where all the primitive changes should progress to the same extent when reaching the transition state \cite{Wilkinson1997,Yang2019}. Given the above definitions, the single TS displayed in Figure \ref{fig:prop}A confirms the concerted other than synchronous feature in the triple proton transfer process. This TS is characterized with a quite positive value of $\delta$(O-H$_\mathrm{a}$)+$\delta$(O-H$_\mathrm{b}$) and a nearly zero value of $\delta$(O-H$_\mathrm{c}$), which indicates that the transfers of H$_\mathrm{a}$ and H$_\mathrm{b}$ have been almost finished, while the transfer of H$_\mathrm{c}$ is still on the half way. It can be concluded that the water autoionization takes place through a concerted and asynchronous proton transfer mechanism. This explains why there was a longer waiting time between the second and third PT event in the previous RETIS study \cite{Moqadam2018}. Interestingly, similar asynchronous PT events were also observed in the alcohol mediated hydroxyquinoline system \cite{Kwon2006,Kang2015}.     
 


Tracking the evolution of each pairwise oxygen distance along the proton transfer coordinate allows a deeper insight into the cooperative relations between proton and oxygen motions. As shown in Figure \ref{fig:prop}B, it is found that the oxygen atoms neighboring to H$_\mathrm{a}$ and H$_\mathrm{b}$ reach their maximal contractions well before the TS (gray line). In contrast, those adjacent to H$_\mathrm{c}$ fulfill the same task much later, which is around the TS point on the reaction path. By comparing Figure \ref{fig:prop}B and \ref{fig:prop}C, it shows clearly that the progress of each proton transfer in the concerted process is controlled well by the neighboring oxygen contraction, and interestingly, corresponding to a similar shortest oxygen separation at about 2.44 \AA. For comparison, in the single jump of hydrated excess proton system, the O-O distance at $\delta$(O-H)=0 was reported to be $\sim$2.46 \AA~\cite{Marx1999}, which implies the extent of pairwise oxygen contraction is insensitive to the PT systems. It is worth mentioning that although the transfer of H$_\mathrm{c}$ is significantly postponed, it still proceeds within a compressed water wire which demonstrates the weak correlation with the preceding two transfers, thus should not be regarded as a stepwise step.

 

There is no doubt that the ion pair at the contact distance, which is separated by two waters (\textbf{ions-2w} for short), is a stable intermediate during water ionization, however, the stability of the more closely contacted \ce{H3O+} and \ce{OH-} ion pair, which is separated by one water (\textbf{ions-1w}) or directly contacted (\textbf{ions-0w}), is yet to be determined. According to the simulation results, neither \textbf{ions-1w} nor \textbf{ions-0w} are metastable structures on the free energy profile. Presumably, both of the two structures are destabilized by the collective compression of the involved oxygen atoms along the water wire (see Figure \ref{fig:prop}B). By comparison, previous AIMD study of water dissociation under uniform electric field observed the formation of \textbf{ions-1w} \cite{Saitta2012}, while another AIMD simulation under extra-high pressure suggested the water dissociation proceeds through the \textbf{ions-0w} state via a bimolecular reaction \cite{Schwegler2001}. External environmental conditions may severely influence the hydrogen bond network in water, such that the stability sequence of these ion pairs might be altered. Moreover, the nuclear quantum effect (NQE) is convinced to further delocalize the protons in the progress of water autoionization \cite{Ceriotti2013,Ceriotti2016,Cassone2020}, raising an open question on the importance of NQE in modeling the dynamics of ionization. However, as we mentioned that the progress of triple proton transfer is controlled mainly by the pace of the water wire contraction which is characterized as the motion of relevant oxygen atoms, the negligible NQE on the much heavier oxygen atoms allows the classical MD simulation to provide a reliable interpretation of the water ionization mechanism.

\begin{figure}[h]
	\centering
	\includegraphics[width=1.0\linewidth] {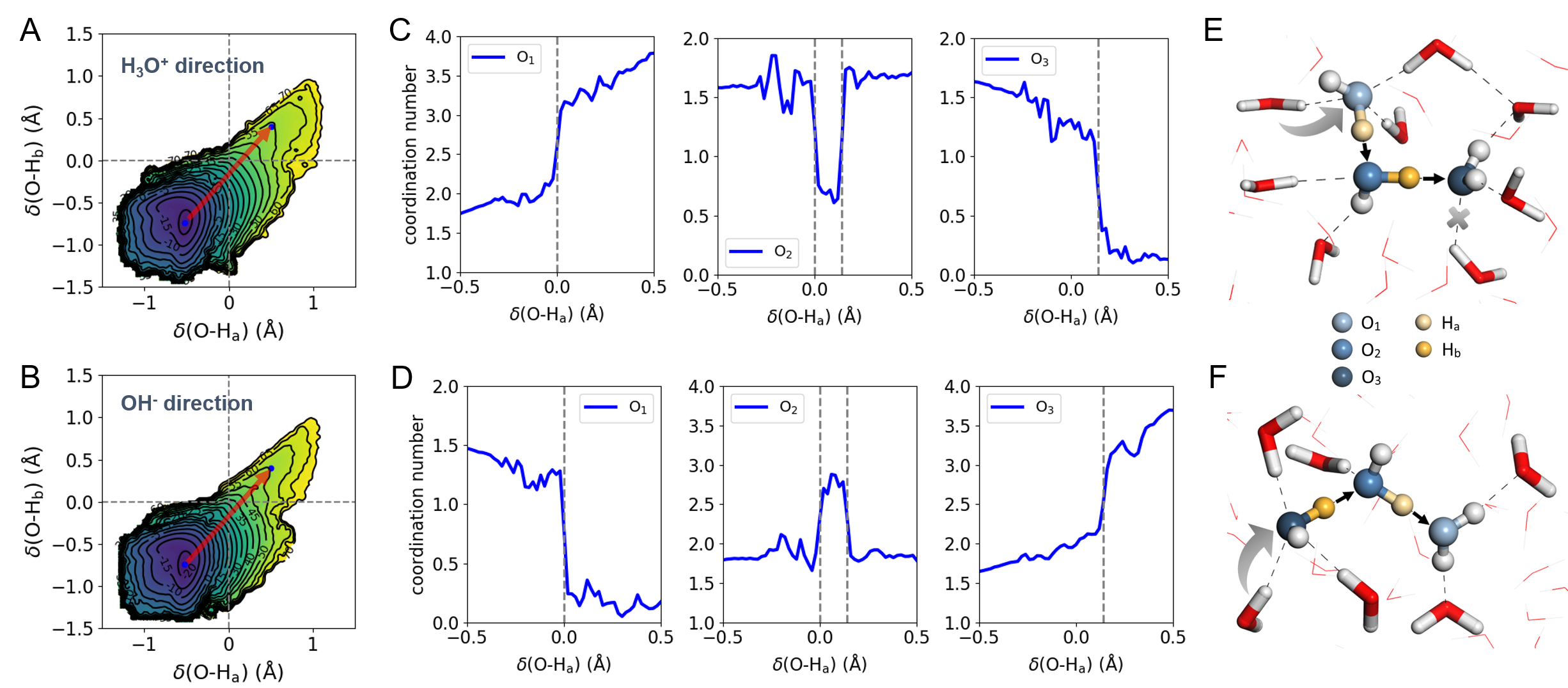}
	\caption{Evolution of coordination number (CN) on oxygen atoms hosting H$_\mathrm{a}$ and H$_\mathrm{b}$ along the proton transfer coordinates. A constraint of $\delta$(O-H$_\mathrm{c}$)$<$0 is applied to filter out the metadynamics configurations belong to the third proton jump stage. Free energy surfaces constructed with $\delta$(O-H$_\text{a}$) and $\delta$(O-H$_\text{b}$) for the PT events along (A) \ce{H3O+} and (B) \ce{OH-} directions, respectively. Red arrowed lines in the two contour plots connecting points (-0.52,-0.74) and (0.5,0.4) characterize an approximate route of double proton transfer. The evolution of CN on the three relevant oxygen atoms as a function of $\delta$(O-H$_\text{a}$) for the transfers of protons H$_\mathrm{a}$ and H$_\mathrm{b}$ along (C) \ce{H3O+} and (D) \ce{OH-} directions. The dashed line marks the point in the transfer route where $\delta$(O-H$_\mathrm{a}$) or $\delta$(O-H$_\mathrm{b}$) equals to zero. Illustrative snapshots of presolvation of the water wires for PT events along (E) \ce{H3O+} and (F) \ce{OH-} directions. }      
	\label{fig:cn}
\end{figure}


\subsection{Dual presolvation and push-pull effect}

The hydrated excess proton was found to preferentially transfer to the neighboring water that possesses the lowest coordination number (CN) in the aqueous solution \cite{Marx1999}, which was later referred to as the mechanism of presolvation. The notion of presolvation can be generalized as the change of HB coordination number on the species ready to accept or donate protons that will initiate the PT events \cite{Tuckerman2002,Berkelbach2009,Marx2010}. Since we are concerned with the concerted triple proton jump during water ionization, presolvation of the four-water wire is investigated here by tracking the number of accepted hydrogen bonds around the involved oxygen atoms along the PT coordinate. We focus attention on the presolvation related to two earlier transfers of H$_\mathrm{a}$ and H$_\mathrm{b}$, since the presolvation of H$_\mathrm{c}$ transfer exhibits obviously the character like a single PT event due to its delayed feature (see Supplementary Fig. S6 for details). Given the reported difference of HB patterns between the \ce{H+} and \ce{OH-} migration processes \cite{Marx1999,Tuckerman2002,Marx2010}, the metadynamics configurations are further classified into the PT events proceeding along \ce{H3O+} and \ce{OH-} directions, respectively. 

As displayed in Figure \ref{fig:cn}A and \ref{fig:cn}B, a concerted transfer route of H$_\mathrm{a}$ and H$_\mathrm{b}$ (red line) is constructed to approximate the minimum free energy path by connecting the initial neutral water state to the artificially assigned \textbf{ion-1w} state. Although determining the location of \textbf{ion-1w} is not unambiguous, the assigned coordinate (0.5, 0.4) can help to well guide the direction where the free energy rises in the slowest way, both in the PT events along \ce{H3O+} and \ce{OH-} directions. The progress of the double proton transfer is found to exhibit relatively high synchronicity, while a 0.14 \AA~delay for $\delta$(O-H$_\mathrm{b}$) can be read from the contours when $\delta$(O-H$_\mathrm{a}$) equals zero. This image reminds us that presolvation of the water wire ready for dissociation may differ from the single proton jump in \ce{H+} or \ce{OH-} transport system \cite{Marx1999,Tuckerman2002}. A series of reweighted CN distributions for each relevant O atom along the PT coordinates (see Supplementary Fig. S5) is computed with the conditional ensemble average algorithm introduced in Methods Section, from where the average CN evolution along the proposed reaction path is extracted. Since the linear transfer path is proportional to $\delta$(O-H$_\mathrm{a}$) and $\delta$(O-H$_\mathrm{b}$), we plot the evolution of CN against $\delta$(O-H$_\mathrm{a}$) for the three specific O atoms, between which H$_\mathrm{a}$ and H$_\mathrm{b}$ are sandwiched, as shown in Figure \ref{fig:cn}C and \ref{fig:cn}D for \ce{H3O+} and \ce{OH-} directions of proton transfer.   

We firstly look into the HB patterns for H$_\mathrm{a}$ and H$_\mathrm{b}$ transfers along the \ce{H3O+} direction. As illustrated from the metadynamics snapshot in Figure \ref{fig:cn}E, two adjacent waters will successively break their covalent bonds, marked with O$_1$-H$_\mathrm{a}$ and O$_2$-H$_\mathrm{b}$, resulting in two newly formed bonds, O$_2$-H$_\mathrm{a}$ and O$_3$-H$_\mathrm{b}$, as well as the ion pair separated by one intermediate water. By inspecting the left panel of Figure \ref{fig:cn}C, the average CN of O$_1$ increases by $\sim$0.4 before $\delta$(O-H$_\mathrm{a}$) reaches zero, which manifests the presolvation phenomenon that a hypercoordinated water is formed prior to the H$_\mathrm{a}$ transfer, as displayed in Figure \ref{fig:cn}E. This observation provides a concrete proof of the hypothesis that a hypercoordinated state is likely to be the nucleation center for water autoionization \cite{Hassanali2011}. The sharp increase from 2.12 to 3.15 of the CN of O$_1$ at gray line arises from the change in affiliation of H$_\mathrm{a}$, on leaving from O$_1$ to O$_2$. As H$_\mathrm{a}$ transfer finishes, the nascent \ce{OH-} is coordinated with nearly four accepted HBs, namely the \ce{OH-}\ce{(H2O)4} complex proposed in the isolated hydrated \ce{OH-} system \cite{Tuckerman2002}. The right panel of Figure \ref{fig:cn}C shows an opposite trend in the CN evolution of O$_3$, with a decrease by $\sim$0.49 before $\delta$(O-H$_\mathrm{b}$) reaches zero. It indicates the necessity of concurrently forming the undercoordinated water environment to trigger the water ionization. Coincidentally, this change of CN is found to be almost the same with the CN decrease ($\sim$0.5) of the proton-receiving water in \ce{H+} diffusion system \cite{Marx1999}. After H$_\mathrm{b}$ transfer, the nascent \ce{H3O+} will exist as a \ce{H9O4+} complex with only three donated HBs coordinated to the ion core, as can be envisaged from Figure \ref{fig:cn}E. Interestingly, the intermediate O$_2$ atom exhibits a ``relay station'' character in the process of double proton transfer. As shown in the middle panel of Figure \ref{fig:cn}C, the CN of O$_2$ is found to essentially unchanged in most range of the transfer path, except in the short period from $\delta$(O-H$_\mathrm{a}$)=0 to $\delta$(O-H$_\mathrm{b}$)=0 (see vertical dashed lines). The sudden dropping followed by a quick restoring of CN is due to the change of the affiliated oxygen atoms of H$_\mathrm{a}$ and H$_\mathrm{b}$, other than a real change of the HB patterns. 

We may rationalize the different behaviors of presolvation on the three oxygen atoms by introducing the Br\o nsted-Lowry acid-base theory \cite{Bronsted1923,Lowry1923}. The increase of CN of O$_1$ indicates the formation of an extra accepted HB on the host water, which will strengthen the acidity of O$_1$ by decreasing its electron density, and hence facilitate the donating of H$_\mathrm{a}$. Similarly, the decrease of CN will increase the alkalinity of O$_3$ and enhance its ability to receive H$_\mathrm{b}$. It seems that the opposite presolvation behaviors on O$_1$ and O$_3$ demonstrate the ``push-pull'' effect that is favorable for water autoionization. In contrast, due to the highly synchronous character of double proton transfer, arriving of H$_\mathrm{a}$ and leaving of H$_\mathrm{b}$ can not benefit simultaneously from the presolvation on O$_2$, e.g., the undercoordination of O$_2$ will increase its alkalinity, which can promote the migration of H$_\mathrm{a}$ but hinder that of H$_\mathrm{b}$. Therefore, the unapparent presolvation on O$_2$ could be interpreted as a ``natural selection'' for resolving such a dilemmatic problem.

 
By comparison, the presolvation of PT events along the \ce{OH-} direction is slightly different from that of the \ce{H3O+} direction. Figure \ref{fig:cn}F displays the dissociation process which starts with the transfer of H$_\mathrm{a}$ from O$_2$ to O$_1$, followed immediately by the transfer of H$_\mathrm{b}$ from O$_3$ to O$_2$. By inspecting Figure \ref{fig:cn}D and \ref{fig:cn}F, O$_1$ and O$_3$ atoms, which are ready to be transformed into \ce{H3O+} and \ce{OH-} ions, show similar presolvation images with the cases along \ce{H3O+} direction. The CN of O$_1$ decreases by $\sim$0.23 before $\delta$(O-H$_\mathrm{a}$) reaches zero, while the CN of O$_3$ increases by $\sim$0.48 before the H$_\mathrm{b}$ transfer moment. Still, the presolvation of middle O$_2$ is found to be negligible. It is approximately threefold coordinated when in the form of the short-life \ce{OH-}, and soon returns to the intermediate water state with two accepted HBs coordinated. After combining the two kinds of PT patterns discussed above, the presolvation of water wire for autoionization is characterized as the simultaneous formation of a pair of hypercoordinated and undercoordinated water molecules that should be separated by one intermediate water. The similar dual presolvation behaviors for \ce{H3O+} and \ce{OH-} directions of double PT events further imply the strong correlations between H$_\mathrm{a}$ and H$_\mathrm{b}$ transfers. Our findings provide some new perspectives on the initiation conditions for the water wire dissociation.

\subsection{Local electric field}

The electric field fluctuations were suggested to play a determinative role in water autoionization \cite{Geissler2001,Geissler2005}. It was later proven that the local solvation environment around the dissociating O-H bond contributes majorly to the internal electric field \cite{Dellago2009}. Here we will explore the correlation between local electric field and presolvation phenomena. By fixing the atomic charges of O and H as -0.8476$e$ and +0.4238$e$ which are taken from the SPC/E rigid water model \cite{Berendsen1987}, the local electric fields on the migrating H$_\mathrm{a}$ and H$_\mathrm{b}$ are estimated roughly with the Coulomb's law, while restricting the contribution from waters within the first solvation shell around the water wire. An additional CV labeled $E$ is defined as the projection of local electric field on the dissociating O-H bond, such that a positive projected electric field is expected to push the protons to dissociation. 

\begin{figure}[h]
	\centering
	\includegraphics[width=1.0\linewidth] {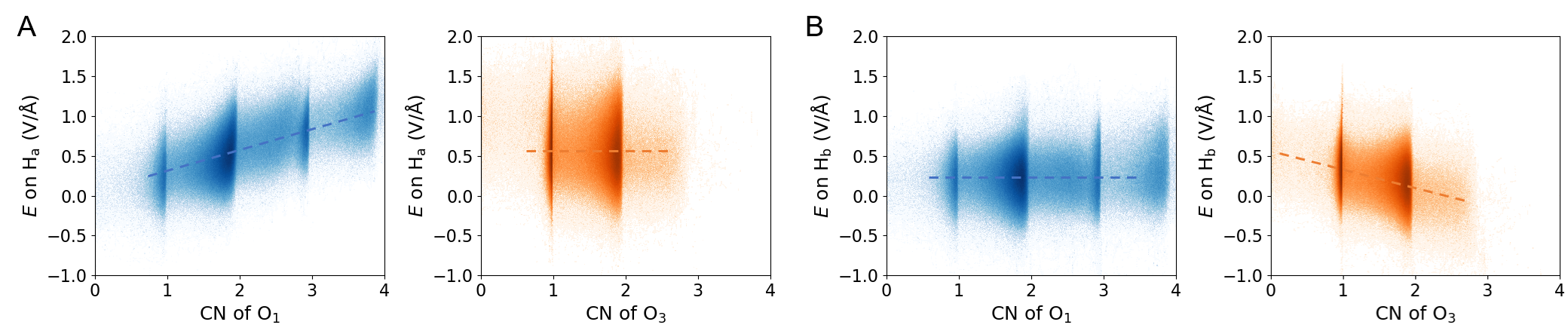}
	\caption{Correlation between local electric field and CN changes in presolvation. The 2D histogram plots in a logarithmic scale of metadynamics configurations described by two CVs, the CN of O$_1$/O$_3$ and the local electric field on (A) H$_\mathrm{a}$ and (B) H$_\mathrm{b}$, for the double PT events along the \ce{H3O+} direction. A constraint of $\delta$(O-H$_\mathrm{c}$)$<$0 is applied in the counting to exclude the configurations belong to the third proton jump stage. The blue and orange colors represent the oxygen atoms which will be transformed to \ce{OH-} and \ce{H3O+}, respectively. All images show a high density of distribution around CN of 2, due to the deep potential well in the neutral water state. Except this feature, the distribution of O$_1$ biases to high CN ($\ge$3) and that of O$_3$ biases to low CN around 1.   }
	\label{fig:ef}
\end{figure}


By projecting the metadynamics configurations into the 2D histogram plots of $E$ and CN of relevant oxygen atoms (O$_1$ and O$_3$), Figure \ref{fig:ef} shows directly the regular patterns of data distributions in the \ce{H3O+} direction of double proton transfer. As the CN of O$_2$ in the intermediate water remains essentially unchanged during the PT events, its effect on the electric field is excluded from the discussion. The larger CN of O$_1$ (next to H$_\mathrm{a}$) contributes mainly to the stronger electric field on H$_\mathrm{a}$, which is, however, much less sensitive to the CN of O$_3$, the oxygen relatively far from H$_\mathrm{a}$. In contrast, the electric field on H$_\mathrm{b}$ shows an increasing trend as the CN of O$_3$ (next to H$_\mathrm{b}$) decreases, but the CN of O$_1$ makes little effect. The results prove that the accepted HB of the dissociating water will pose a repulsive electrostatic force on the proton H$_\mathrm{a}$ to facilitate its leaving, as corresponding to the push effect; while the breakage of accepted HB on the proton-receiving water will reduce such kind of repulsion, and in turn attract the arrival of proton H$_\mathrm{b}$, namely the pull effect. Indeed, part of these findings is in accord with previous simulation which demonstrated that a larger number of accepted HB could enhance the local electric field on the stretching O-H bond \cite{Dellago2009}. Consequently, both the hypercoordinated state and the undercoordinated state in dual-presolvation mechanism are suggested to be correlated with the strong local electric field on the nearest migrating protons, and a similar conclusion holds for the \ce{OH-} direction of double proton transfer (see Supplementary Fig. S7 for details). These findings could build connections between presolvation and electric field fluctuations in triggering the water ionization.

\section{Conclusions}

We investigate the mechanism of triple proton jump in water autoionization to form the \ce{H3O+}\ce{\bond{...}}\ce{OH-} pair that separated in the contact distance at room temperature, with the nanosecond-scale metadynamics simulations based on DFT-parameterized atomic neural network potentials to ensure a reliable sampling of the dissociation path. The single transition state on the dissociation free energy surface is characterized as a stage where the transfers of H$_\mathrm{a}$ and H$_\mathrm{b}$ are almost finished while the transfer of H$_\mathrm{c}$ is on the half way. This image establishes that the hydrogen-bonded water wire goes through an asynchronous and concerted transfer of three protons. By comparing the evolution of $\delta$(O-H) and pairwise oxygen distance, the progress of each proton transfer is found to be well controlled by the neighboring oxygen contraction, implying the neglect of nuclear quantum effect will not alter the essential conclusions on the ionization mechanism. Based on critical points in the free energy profiles, the calculated water equilibrium constant 0.724$\times 10^{-14}$ is found to be quantitatively consistent with experimental value, while the overall rate constants for water dissociation and ions recombination are computed as 1.566$\times 10^{-3}$ s$^{-1}$ and 0.216 ps$^{-1}$ from steady-state approximation, also in good agreement with the experimental measurement. After detailed analysis of the CN evolution on oxygen atoms hosting H$_\mathrm{a}$ and H$_\mathrm{b}$, the dual-presolvation mechanism is proposed and characterized as the simultaneous formation of a pair of hypercoordinated and undercoordinated water molecules that separated by one intermediate water. The 2D density distributions of metadynamics configurations reveal that the CN changes in the presolvation phenomena are correlated with the strong local electric field, especially for the HB patterns nearest to the dissociating protons, both showing the same push-pull effect on driving the water ionization. In summary, we advance the understandings of water autoionization from concerted to asynchronous, from exclusive presolvation to dual presolvation, as well as the contribution to internal electric field. 

The successful combination of biased dynamics and neural network potentials has provided a demonstrative theoretical research on the multiscale condensed matter system, which displays fully its high efficiency and reliability in studying complex chemical systems. By utilizing designed CVs to describe water dissociation along the water wire, our work elucidates the nature of triple proton transfer and helps to comprehend the solvent environment in facilitating the ionization process. We anticipate these new understandings of water characteristics along with our theoretical scheme can be referenced in the study of multiple proton transfer in biological systems, or the water dissociation under high pressure, within electric double layer, as well as in other external conditions. 


\section{Methods}

\subsection{DFT calculations}

All the DFT calculations were carried out in VASP 5.4.4 package \cite{Kress1993,Kress1996}. The wave functions of valence electrons were expanded in the plane wave basis set with a energy cutoff of 400 eV. Core electrons were described with pseupotentials using projector augmented wave (PAW) method \cite{Bloch1994}. RPBE functional formulated with generalized gradient approximation (GGA) \cite{Perdew1992,Hammer1999} was used to calculate the exchange and correlation energy. DFT-D3/zero correction method was applied to include the van der Waals (vdW) energy \cite{Grimme2010}. RPBE-D3/zero functional has been proved to be an appropriate choice to describe the liquid water properties \cite{Grob2016,Morawietz2016}. The Brillouin zone was sampled by a single $\Gamma$ point. 

\subsection{AIMD simulations}
Bulk water system was constructed by a $12.42\times12.42\times12.42$ \AA$^3$ cubic box containing 64 randomly distributed water molecules at a density of 1 g/cm$^3$. AIMD simulations were conducted with 1 fs time step in the canonical ensemble (NVT) at temperature of 298 K. The convergence criterion for DFT energy and wave function was set to 10$^{-6}$ eV. Additional AIMD simulations at the range of 400-800 K were performed to improve the diversity of water samples. The ionic state was simulated and sampled by artificially separating one pair of \ce{H3O+} and \ce{OH-} in the above water box with distances ranging from 2.7 to 8.6 \AA~in the constrained molecular dynamics \cite{Ryckaert1977}, where the number of covalent bonds of center oxygen atoms was constrained to avoid quick recombination of ions. The constrains were later removed to sample the configurations of recombination process with standard AIMD.    

\subsection{Training NNPs}
The atomic neural network potentials (NNPs) were trained using n2p2 package \cite{Singraber2021}. All NNPs consist of a set of feed-forward neural networks with two hidden layers each containing 25 nodes. The local chemical environment of H and O atoms were described by a total of 27 and 30 symmetry functions (SFs, including radial and angular types) introduced in a previous study \cite{Morawietz2016}, respectively, each with a 6.0 \AA~cutoff radius. Initial reference water samples contain 1939 configurations randomly selected from AIMD trajectories. Then the preliminary NNPs were constructed and employed in NVT MD calculations with the version of NNP library based LAMMPS program \cite{Singraber2019,Aidan2022}. The structures with extrapolation warnings reported in MD were filtered by the vector distance criterion, which can be expressed as
\begin{equation}
	R^2_\mathrm{vd} = \min\left[(\boldsymbol{G} - \boldsymbol{G}_\mathrm{ref,1})^2,(\boldsymbol{G} - \boldsymbol{G}_\mathrm{ref,2})^2,...,(\boldsymbol{G} - \boldsymbol{G}_{\mathrm{ref},n})^2\right]
\end{equation}

\noindent For a selected configuration with a set of normalized SFs ($\boldsymbol{G}$), if the calculated $R_\mathrm{vd}$ is larger than our defined cutoff distance $R_\mathrm{c}$, this configuration will be accepted as a new reference data and recomputed by DFT. $\boldsymbol{G}_{\mathrm{ref},i}$ denotes the normalized SFs of the $i$-th structure in the reference data set, while the total number $n$ will increase along with the screening process. $R_\mathrm{c}$ was artificially set according to the expected number of newly added reference structures. Several cycles from NNPs training to configuration selection were performed until the NNPs are ready for the ns-scale MD simulations with few extrapolation warning. Final reference data set for liquid water contains 4498 configurations, and the NNPs based MD simulation can reproduce the radial distribution function and self-diffusion coefficient well as compared with the experimental data (see Supplementary Notes and Fig. S1 for details). The root mean squared errors (RMSE) for energy and force of reference data set were calculated to be 0.178 meV/atom and 43.535 meV/\AA, respectively.   

In the construction of NNPs for water dissociation, 1271 configurations of ion pair and its recombination process were extracted from AIMD trajectories and added into the above water samples. The refinement of NNPs was achieved by supplementing structures from metadynamics along the chosen collective variables that drive the dissociation of water molecules. The vector distance criterion was again employed to filter configurations that distributed within different regions in the entire CV space, which was artificially divided into regions for pure water, ion pair and reaction process. Such division favors an uniform sampling of different reaction stages, since most configurations are located in the deep potential well of neutral water state. The final reference data set for water dissociation contains 10212 configurations, and the trained NNPs can support stable metadynamics simulations over 10 ns. The RMSE for energy and force of reference data set were calculated to be 0.335 meV/atom and 49.625 meV/\AA, respectively.        

\subsection{Metadynamics and CVs}
NNPs-based metadynamics simulations were performed using LAMMPS code interfaced with n2p2 and the metadynamics plugin engine PLUMED \cite{Gareth2014}. We applied well-tempered metadynamics \cite{Parrinello2008} at 298 K with deposition stride of 0.2 ps, starting hill height of 1.0 kJ/mol, Gaussian widths of 0.2 and 1.2 \AA~for two CVs, and bias factor of 9. Another two sets of parameters were adopted to examine the consistence of free energy surfaces as introduced in Supplementary Notes. All the metadynamics simulations were initiated from a periodic cubic box containing 64 \ce{H2O} molecules (corresponding to $\sim$0.869 M for ions), with a time step of 0.5 fs and a total simulation time of 10 ns.     

The two collective variables controlling the progress of water dissociation are expressed as follows,
\begin{equation}\label{eq:cv1}
	\mathrm{cv1}=\sum_{i=1}^{N_\mathrm{O}}[n_i(\boldsymbol{R})-2]^2 	\text{~~~where,~}n_i(\boldsymbol{R})=\sum_{j=1}^{N_\mathrm{H}} \frac{1-(R_{ij}/R_0)^{16}}{1-(R_{ij}/R_0)^{56}}
\end{equation}
\begin{equation}\label{eq:cv2}
	\mathrm{cv2}=-\sum_{i,k>i}^{N_\mathrm{O}} q_i q_k R_{ik} \text{~~~where,~}q_i=\sum_{j=1}^{N_\mathrm{H}} \frac{e^{-\lambda R_{ij}}}{\sum_{i'=1}^{N_\mathrm{O}}e^{-\lambda R_{i'j}}}-2
\end{equation}
\noindent cv1 denotes the number of ions, where $n_i$ corresponds to the number of covalent bonds for $i$-th O atom. $R_{ij}$ is the distance between O$_i$ and H$_j$, and cutoff $R_0$ was set as 1.32 \AA~\cite{Hassanali2011}. $N_\mathrm{O}$ and $N_\mathrm{H}$ denote the total number of oxygen and hydrogen atoms in the simulated box. cv2 is defined as the distance between ionic species which was proposed by Grifoni et al.~\cite{Grifoni2019}. $R_{ik}$ denotes the distance between O$_i$ and O$_k$. $q_{i}$ refers to the electric charge of the chemical species possessing O$_i$. With $\lambda$ set as 8 \AA$^{-1}$, the value of $q_{i}$ ranges from roughly -1 (\ce{OH-}), 0 (\ce{H2O}) to +1 (\ce{H3O+}). Obviously, cv1 and cv2 satisfy the conditions of continuous and differentiable that required in conducting metadynamics. Besides, the two CVs could drive the water dissociation and the formation of ion pair efficiently, enabling an abundant sampling along the reaction process. A restraining potential expressed as $200(\mathrm{cv1}-2)^2+8(\mathrm{cv2}-10)^2$ was applied in the region of cv1>2 or cv2>10 \AA~to avoid two independent dissociation events occurring simultaneously. 
The coordination number of accepted HBs of the $i$-th oxygen atom is expressed as
\begin{equation}\label{eq:cn}
	\mathrm{CN}_i = \sum_{k=1}\left[\frac{1-(R_{ik}/R_1)^{16}}{1-(R_{ik}/R_1)^{56}}\right] \cdot \left[\frac{1-(\theta_{ik}/\theta_1)^{16}}{1-(\theta_{ik}/\theta_1)^{56}} \right]
\end{equation} 
where the cutoff $R_1$ was set as 3.3 \AA, and $\theta_1$ was set as $\pi/4$. $\theta_{ik}$ refers to the supplementary angle of the O$_k$-H-O$_i$ hydrogen bond. The summation was performed with the restriction that O$_k$ and O$_i$ should be the nearest and next-nearest oxygen neighbors of the H atom in between. 

Note that CVs used to analysis the concerned properties and describe the free energy surfaces can be non-differentiable. The free energies were then calculated with the reweighting scheme introduced in the work of Tiwary and coworker \cite{Tiwary2015}. The conditional ensemble average of an arbitrary position-dependent operator $O(\boldsymbol{R})$, given that the set of CVs is equal to $\boldsymbol{s}$, was calculated by the following formula,
\begin{equation}
	\langle O\rangle_{\boldsymbol{s}}=\frac{\sum_{t'=0}^t w_{t'} K(\boldsymbol{s}_{t'}-\boldsymbol{s})O_{t'}}{\sum_{t'=0}^t w_{t'} K(\boldsymbol{s}_{t'}-\boldsymbol{s})}
\end{equation}    
\noindent where $w_{t'}$ refers to the calculated weight at time $t'$. $K(\boldsymbol{s}_{t'}-\boldsymbol{s})$ denotes a Gaussian kernel function centered at the current value $\boldsymbol{s}_{t'}$. The corresponding Gaussian widths for CVs were set as 0.04 \AA~and 0.01 \AA~in the $R$(O-O) and CN calculations, respectively. Another module to calculate the ensemble average of a specific variable, e.g. $R_\mathrm{cd}$, with restriction applied on CVs, is expressed as
\begin{equation}
	\langle R_\mathrm{cd}\rangle =\frac{\sum_{t'=0,\boldsymbol{s}_{t'} \in Z}^t w_{t'} R_{\mathrm{cd},t'}}{\sum_{t'=0,\boldsymbol{s}_{t'} \in Z}^t w_{t'}}
\end{equation} 
\noindent The restriction condition of averaging the contact distance was set as $\sum_3\delta$(O-H)>1.38 \AA, ensuring the sampled ions are separated by two waters. The two modules for ensemble average calculation along with the construction of all the CVs were implemented with an in-house extended version of PLUMED code.

\section{Acknowledgments}

The authors thank Profs. Yi Gao and Ye Mei for constructive suggestions on the manuscript. Financial support from the China NSF (Grand No. 22073041) is gratefully acknowledged. We thank the High Performance Computing Center of Nanjing University for computational resources.

\bibliography{ref}
\bibliographystyle{pnas-new}

\end{document}